\documentstyle[aaspp4,emulateapj5,apjfonts]{article}

\lefthead{van der Wel et al.}
\righthead{The FP at z=1 for Field Galaxies}
\slugcomment{Accepted for publication in Astrophysical Journal Letters}

\begin{document}

\title{The Fundamental Plane of Field Early Type Galaxies at $z=1$\altaffilmark{1}}

\author{A.~van~der~Wel\altaffilmark{2}, M.~Franx\altaffilmark{2}, P.G.~van~Dokkum\altaffilmark{3}, and H.-W.~Rix\altaffilmark{4}}
\altaffiltext{1}{Based on observations collected at the European Southern Observatory, Chile (169.A-0458).}
\altaffiltext{2}{Leiden Observatory, P.O.Box 9513, NL-2300 AA, Leiden, The Netherlands}
\altaffiltext{3}{Yale University, New Haven, CT 06520-8101}
\altaffiltext{4}{Max-Planck-Institut f\"ur Astronomie, K\"onigstuhl 17, D-69117 Heidelberg, Germany}
%
%
%

\begin{abstract}

We present deep VLT spectra of early type galaxies at $z \approx 1$ in
the Chandra Deep Field South, from which we derive velocity
dispersions.  Together with structural parameters from Hubble Space
Telescope imaging, we can study the Fundamental Plane for field early
type galaxies at that epoch. We determine accurate mass-to-light
ratios and colors for four field early type galaxies in the redshift
range $0.96<z<1.14$, and two with $0.65<z<0.70$.

The galaxies were selected by color and morphology, and have generally
red colors. Their velocity dispersions show, however, that they have a 
considerable spread in
mass-to-light ratios (factor of 3). We find that the colors and
directly measured mass-to-light ratios correlate well, demonstrating
that the spread in mass-to-light ratios is real and reflects
variations in stellar populations.

The most massive galaxies have mass-to-light ratios comparable to
massive cluster galaxies at similar redshift, and therefore have stellar
populations which formed at high redshift ($z>2$). The lower mass
galaxies at $z \approx 1$ have a lower average mass-to-light ratio,
and one is a genuine 'E+A' galaxy. The mass-to-light ratios indicate 
that their luminosity weighted ages are a factor of three younger at the epoch 
of observation, due to either a late formation redshift, or due to 
late bursts of star formation
contributing 20 - 30\% of the mass.

\end{abstract}


\section{Introduction}

The formation and evolution of early type galaxies is one of the
major challenges for current structure formation models.
Models of hierarchical structure generally predict that field early
type galaxies form relatively late 
({e.g.,} {Diaferio} {et~al.} 2001).

One of the prime diagnostics of the formation history of early type
galaxies is the evolution of the mass-to-light ratio as measured
from the Fundamental Plane
({Franx} 1993).

Studies of the evolution of the luminosity function together with the
evolution of $M/L$ quantifies the evolution of the
mass function. 

Previous studies of the evolution of mass-to-light ratios have
produced consistent results for the evolution of massive cluster early type
galaxies:
the evolution is slow, consistent with star formation redshifts $z\approx 2$ 
({e.g.,} {van Dokkum} \& {Stanford} 2003).

On the other hand, studies of the evolution of field galaxies have
yielded more contradictory results: whereas early studies produced
slow evolution 
({e.g.,} {van Dokkum} {et~al.} 2001; {Treu} {et~al.} 2002; {Kochanek} {et~al.} 2000),
more recently evidence for much faster evolution was found
by 
{Treu} {et~al.} (2002)
and 
{Gebhardt} {et~al.} (2003),
whereas other authors found
that the majority of field early types evolve slowly, with a 
relatively small fraction of fast evolving galaxies 
({e.g.,} {Rusin} {et~al.} 2003; {van Dokkum} \& {Ellis} 2003; {van de Ven}, {van Dokkum}, \& {Franx} 2003; {Bell} {et al.} 2003).

These previous measurements suffered from several uncertainties:
The signal-to-noise ratios of the spectra were generally quite low,
much lower than usual for nearby studies of the FP
({e.g.,} {Faber} {et~al.} 1989; {J\o rgensen}, {Franx}, \& {Kj\ae rgaard} 1996).

Those studies based on lensing galaxies used stellar velocity dispersions
derived from image separations.

In this \textit{Letter}, we present high signal-to-noise spectra
and accurate measurements of the mass-to-light ratios of four
elliptical field galaxies around redshift one and up to $z=1.14$ and
two at $z\sim0.7$ in the Chandra Deep Field South (CDFS). 
The signal-to-noise ratios are comparable to those obtained for nearby 
galaxies.
Together
with accurate multi-band photometry available for the CDFS,
we have measured the accurate mass-to-light ratios and rest frame
optical colors
at $z\sim1$.

Throughout this \textit{Letter} we use Vega magnitudes, and assume a
$\Lambda$-dominated cosmology
($(\Omega_M,\Omega_{\Lambda})=(0.3,0.7)$), with a Hubble constant of
$H_0=70~km~s^{-1}~Mpc^{-1}$.

\begin{figure*}
\leavevmode
\hbox{%
\epsfxsize=10.5cm
\epsffile{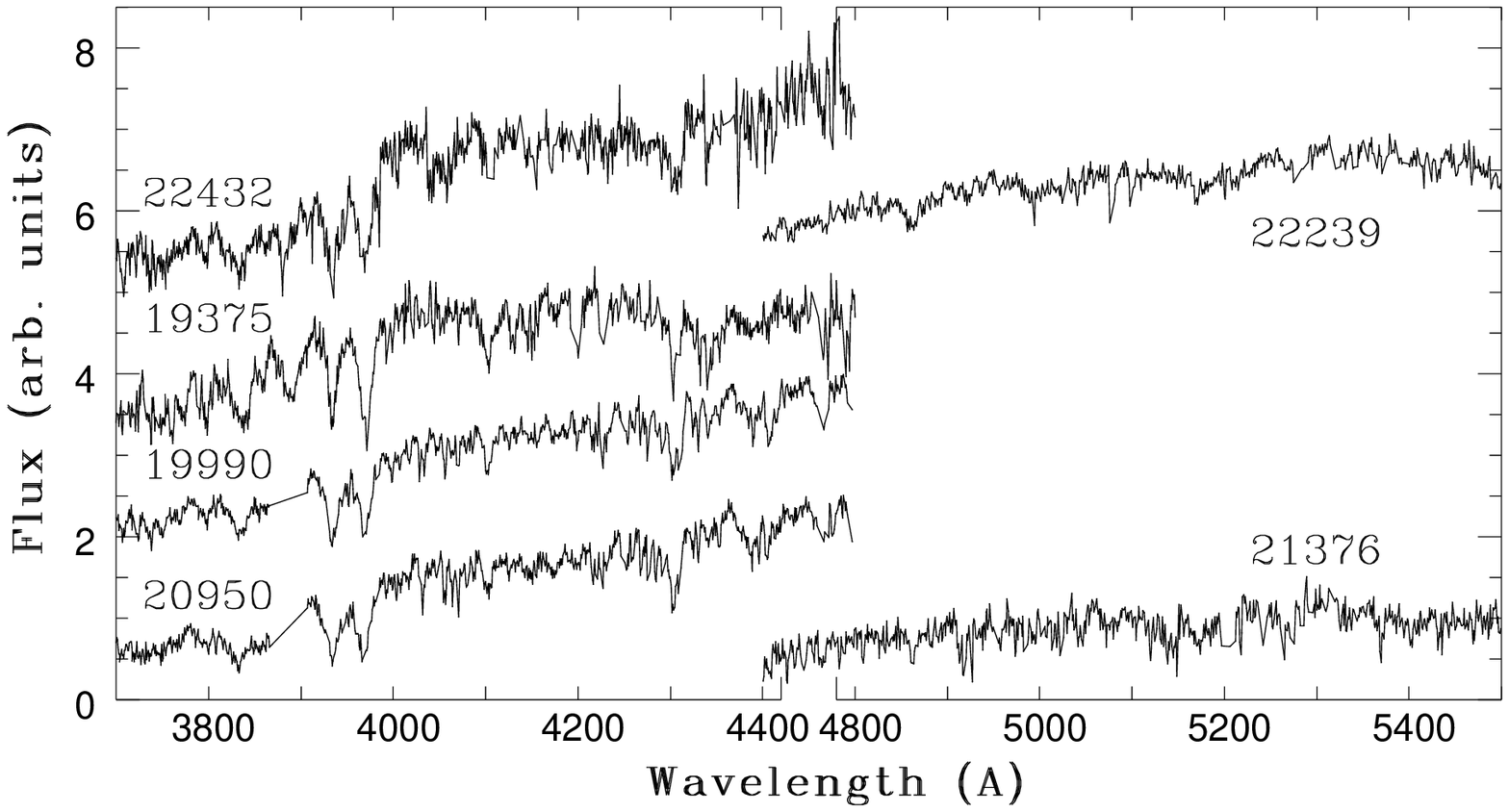}
\epsfxsize=6.35cm
\epsffile{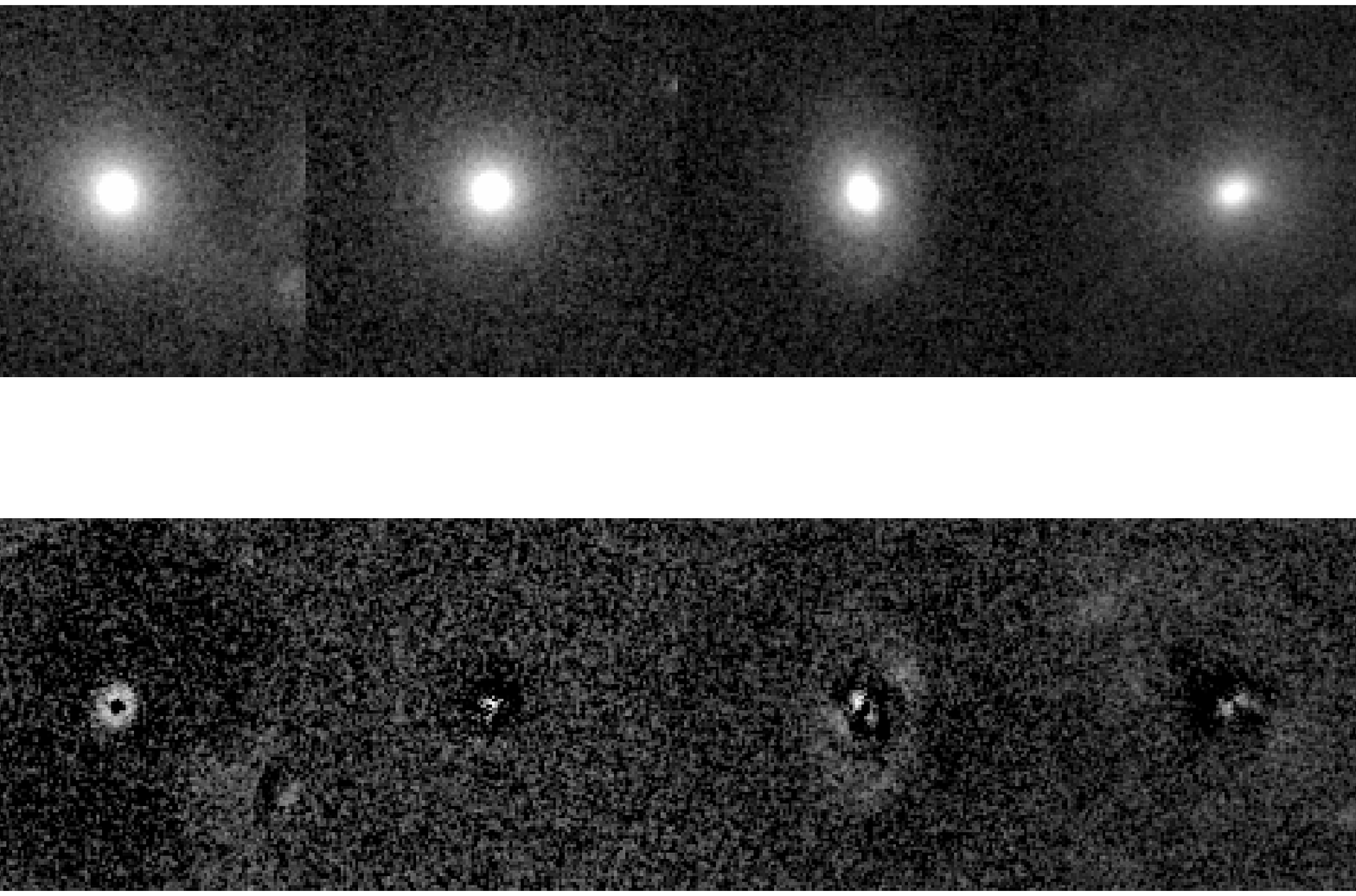}}
\figcaption{
\small
Left panel: unsmoothed restframe spectra of the siz objects with velocity dispersions. Regions with bright sky lines are interpolated. The wavelength scale is interrupted at $\lambda=4500$\AA.
Right panel: ACS images (F850LP) of the four galaxies at $z\sim1$ and the residual images from the $r^{1/4}$-fit. From left to right: 20950, 19990, 19375, 22432.
\label{fig:spec}
\label{fig:im}}
\end{figure*}

\section{Spectroscopy}
\subsection{Sample selection and Observations}\label{sec:obs}

The galaxies were selected from the COMBO17 catalogue
(see {Wolf} {et al.} 2003),
and imaging obtained by the 
Great Observatories Origin Deep Survey
(GOODS\footnote{http://www.stsci.edu/science/goods/}, data release v0.5)
from the
Advanced Camera for Surveys (ACS) on the \emph{Hubble Space Telescope}.
We selected compact, regularly shaped galaxies with photometric redshifts higher than 0.8 and
$I-z \ge (I-z)_{Sbc},~z< 22$. $(I-z)_{Sbc}$ denotes the color of the Sbc template of 
({Coleman}, {Wu}, \& {Weedman} 1980)
at the photometric redshift. This template has $(U-V)_{z=0}=0.95$. 
The typical uncertainty in the color is 0.1 mag.
Lower priority galaxies were included with either lower redshifts 
or later types. 

The CDFS was observed in MXU-mode with the Focal Reducer/Low
Dispersion Spectrograph 2 (FORS2) on the Very Large Telescope (VLT)
Unit Telescope 4 during 3 runs from 2002 September through 2003 Februari,
for  a total of 14 hours.
The 600z grism (central wavelength
9010\AA, resolution 5.1\AA~ or $\sigma_{instr}=72~km~s^{-1}$) was used.
During the observations towards the CDFS the seeing varied between
$0\farcs7$ and $1\farcs5$, with a median seeing of about $1''$. The
sky was clear all the time.

\subsection{Velocity Dispersions}

It turned out that 10 out of the 11 high priority objects at $z_{phot}\sim1$ 
have the spectrum of a quiescent galaxy, the other one 
has a bright [\ion{O}{2}] emission line. The brightest four 
($z\lesssim21.0$, independent of the color) had sufficient 
signal-to-noise ratios to perform reliable dispersion measurements 
(see Table \ref{t:t1}).
The restframe spectra of these 4 galaxies are shown in Figure
\ref{fig:spec}. They all show a strong 4000\AA-break and
Ca-lines. Balmer lines (especially the $H\delta$-line) are also
present, though varying in strength from object to object (see Table
\ref{t:t1}). Object 19375 is an 'E+A' galaxy, according to the
criteria used by 
{Fisher} {et~al.} (1998).
For 2 ellipticals with $0.65<z<0.70$ we also have sufficient signal to determine a velocity dispersion.

Dispersions were measured by convolving a template star spectrum to
fit the galaxy spectrum as outlined by 
{van Dokkum} \& {Franx} (1996).
We tested this procedure extensively, using different
template stars and masking various spectral regions. The final values
(see Table \ref{t:t1})
for the velocity dispersions were obtained by masking the Ca H and K
and Balmer lines and using the best fitting template spectrum,
which was  a high-resolution solar model
spectrum\footnote{http://bass2000.obspm.fr} 
smoothed and
rebinned to match the resolution of the galaxy spectra. The Ca-lines
were not included in the fit because this greatly reduced the dependence
of the measured velocity dispersions on template type.
The tests using different templates and
different masking of the Ca lines indicate that the systematic
uncertainty
is $\sim5\%$. 

In order for the results to be comparable to previous studies, an 
aperture correction as described by 
{J\o rgensen}, {Franx}, \& {Kj\ae rgaard} (1995b)
was applied to obtain velocity dispersions within a circular aperture 
with a radius of $1\farcs 7$ at the distance of the Coma cluster. This correction is $\sim7\%$.

This is the first extensive sample of such objects at $z>0.9$ with high $S/N$.
({see} {van Dokkum} \& {Ellis} 2003; {Treu} {et~al.} 2002; {Gebhardt} {et~al.} 2003 {for other spectroscopic studies}).

\begin{figure*}
\begin{center}
\leavevmode
\hbox{%
\epsfxsize=7.8cm
\epsffile{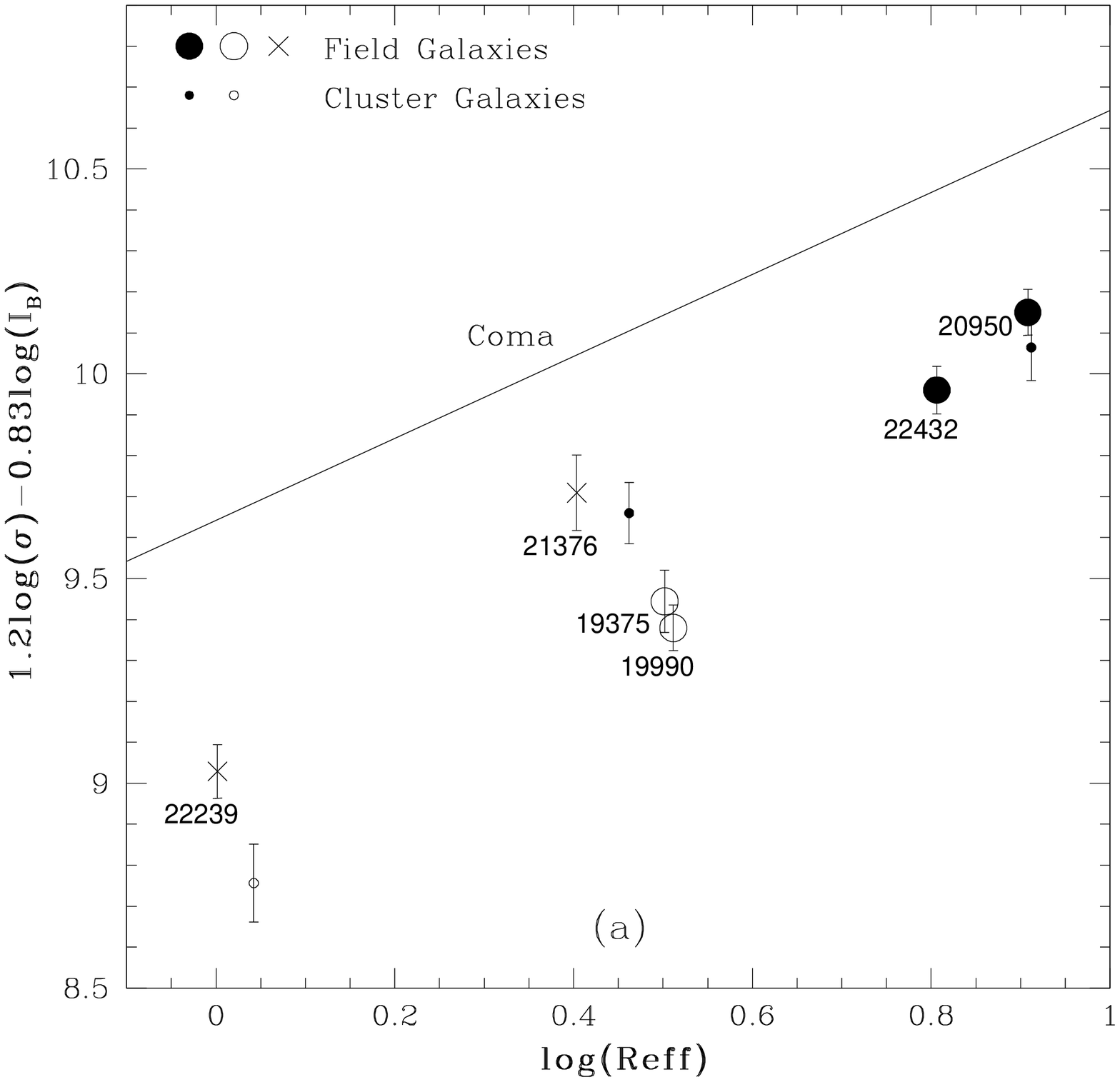}
\epsfxsize=7.8cm
\epsffile{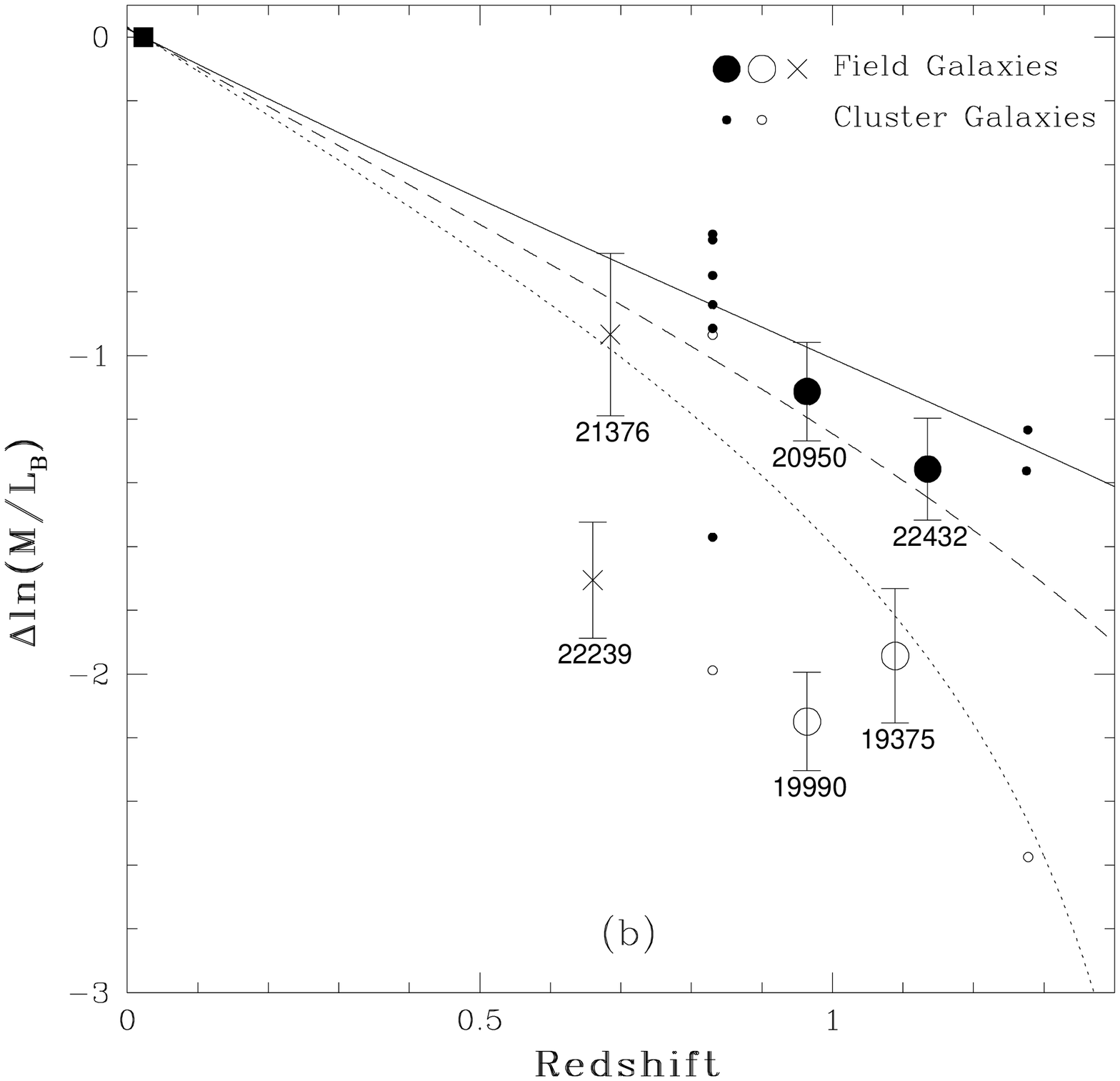}}
\figcaption{
\small
Figure \ref{fig:FP}a shows the FP points of the field galaxies presented in this paper (errors include a 5\% systematic effect in the velocity dispersions), the FP points of cluster galaxies at $z=1.27$ (van Dokkum \& Standford 2003), and the FP of the Coma cluster (J\o rgensen et al. 1996). Offsets in $M/L_B$ from the Coma cluster FP (the square, derived from J\o rgensen et al. 1996) are shown in Figure \ref{fig:FP}b. Besides the $z=1.27$-cluster, this figure also contains the data from van Dokkum et al. (1998) on the MS1054 cluster at $z=0.83$. The full, dashed and dotted curves are the model predictions for a single burst of starformation with a Salpeter IMF for redshifts 3, 2, and 1.5, respectively. Filled symbols indicate galaxies with masses $M>3\times10^{11}M_{\odot}$, other symbols indicate galaxies less massive than that. Crosses and squares distinguish between galaxies at $z<0.8$ and $z>0.8$, respectively. All galaxies occupying the region below the $z=1.5$ model curve are 'E+A' galaxies, except 19990, and have masses less than $3\times10^{11}M_{\odot}$.  The more massive galaxies have significantly older stellar populations.
\label{fig:FP}}
\end{center}
\end{figure*}

\section{Photometry}\label{sec:phot}

Photometry and structural parameters were determined from the GOODS
ACS images (data release v1.0). Images are available in 4 filters (F435W, F606W, F775W,
F850LP), which we refer to as $b$, $v$, $i$, and $z$, respectively.

For each object, the effective radius ($r_e$) and the surface
brightness at the effective radius ($\mu_e$) were obtained by fitting
an $r^{1/4}$-profile, convolved by the PSF 
({van Dokkum} \& {Franx} 1996).
$z$-band
images were used for the $z\sim1$ objects and $i$-band images for the
$z\sim0.7$ objects. Stars were used as the PSF. The resulting values
for $r_e$ and $\mu_e$ vary by $\approx10\%$ when using different stars, 
but also correlate such that the error is almost parallel to the FP
({van Dokkum} \& {Franx} 1996).
Therefore, the errors in our results are dominated by the
errors in the velocity dispersions. The results are listed in Table
\ref{t:t1}. The images of the $z\sim1$ objects and the residuals of the fits
are shown in Figure \ref{fig:im}.


To determine the $i-z$ and $v-i$ colors, fluxes were 
calculated from the $r^{1/4}$-model  
within the measured effective radius. 
To this model flux we add the flux within the same radius in the residual 
images.
We corrected for galactic extinction based on the
extinction maps from 
{Schlegel}, {Finkbeiner}, \& {Davis} (1998).
The correction is extremely small: $E(B-V)=0.007$.

Restframe $B$-band surface brightnesses and restframe 
$U-V$ colors were obtained by transforming
observed flux densities in two filters to a restframe flux density exactly as
outlined by 
{van Dokkum} \& {Franx} (1996).
The spectral energy
distribution used to calculate the transformations is the early type
spectrum from 
{Coleman} {et~al.} (1980).
We found the same results for other template spectra.
The results are listed in Table \ref{t:t1}.

\section{Mass-to-Light Ratios from the FP}

Figure \ref{fig:FP}a shows the FP for the 6 field
galaxies described above, and the FP for Coma derived
by 
{J\o rgensen} {et~al.} (1996).
Additionally, we show the results from
{van Dokkum} \& {Stanford} (2003) 
on three cluster galaxies at $z=1.27$. The
offsets of the high redshift galaxies from the Coma FP are a measure
of the evolution of $M/L$.
We show the evolution of
$M/L$ in Figure \ref{fig:FP}b as a function of
redshift.

Obviously, the field galaxies at $z>0.6$ span a wide range in
offsets, approximately a factor of 3 in $M/L$. 
The errorbars on the
individual points are much smaller than the offsets. A model with a
single formation redshift can be ruled out at the 99\% confidence
level,
as measured from the $\chi ^2$-method.
The restframe colors of the galaxies confirm the reality of the
variations in the mass-to-light ratios. As shown in Figure \ref{fig:UVML}, a
very strong correlation exists between the colors and the
mass-to-light
ratios in the direction expected from population sysnthesis models.

The good correlation 
demonstrates that colors can be used to estimate the mass-to-light
ratios, as applied, for example, by 
{Bell} {et~al.} (2003)
to a large
sample of field early type galaxies.

Note that galaxies in our study lie fairly close to the red
sequence, and were characterised by 
{Bell} {et~al.} (2003)
to have
red colors. The overall spread in colors of field galaxies is much
larger (1.5 mag) compared to the spread found here (0.3 mag).

\section{Discussion}

On the basis of our high signal-to-noise spectra we have found a
rather
wide range in $M/L$ for early type galaxies
at $z=1$, indicating a range in star formation histories. The
 mass-to-light ratios and colors are well correlated, as predicted
by stellar population models.
Hence the scatter in $M/L$ is real.

The results agree surprisingly well with earlier results based on
lensing galaxies. 
{Rusin} {et~al.} (2003)
and 
{van de Ven} {et~al.} (2003)
found
a range in $M/L$, and 
{van de Ven} {et~al.} (2003)
found a similar
correlation between restframe colors, and $M/L$.

Other authors  found either 
low mass-to-light ratios 
({e.g.,} {Treu} {et~al.} 2002; {Gebhardt} {et~al.} 2003),
or high mass-to-light ratios 
({e.g.,} {van Dokkum} \& {Ellis} 2003),
and this is most likely due to (still unexplained) sample selection effects.
The last authors found that galaxies with residuals from the $r^{1/4}$
profile had young ages. However, we find no such relation in our sample.

\null
\vbox{
\begin{center}
\leavevmode
\hbox{%
\epsfxsize=7.8cm
\epsffile{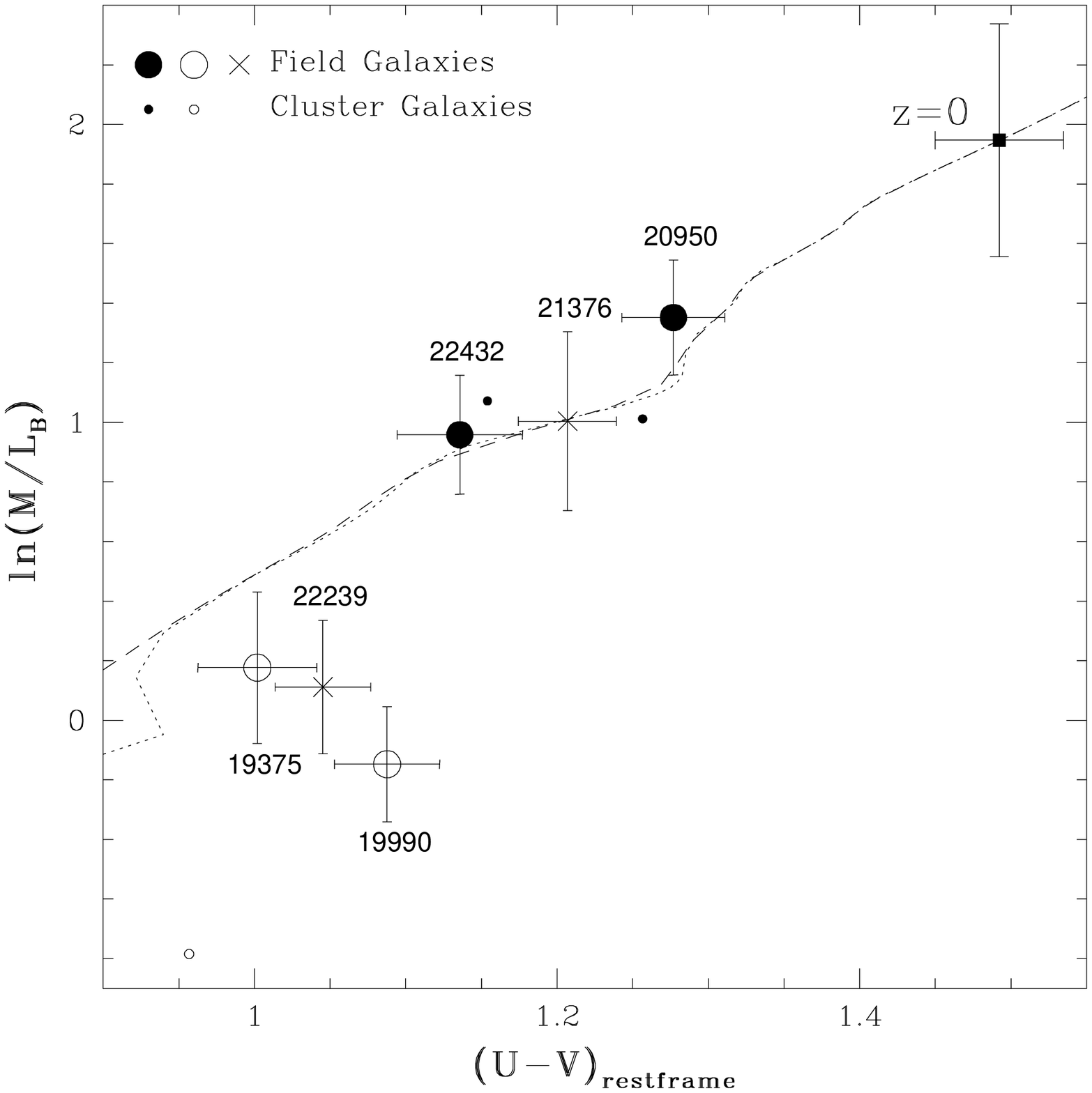}}
\figcaption{\small
Restframe $U-V$ color versus $M/L_B$ in solar units. Filled symbols are objects more massive than $M>3\times10^{11}M_{\odot}$, open symbols represent the less massive ones. The cluster galaxies are the $z=1.27$-galaxies from Figure \ref{fig:FP}. The redshift zero datapoint is the average for galaxies with $\sigma>150~km~s^{-1}$ in the clusters Abell 194 and DC2345-28 (J\o rgensen, Franx, \& Kj\ae rgaard 1995a). The lines are solar metallicity Bruzual \& Charlot (2003) models with constant star formation during the first $200~Myr$ (dotted) and exponentially decaying star formation on the same time scale (dashed).
\label{fig:UVML}}
\end{center}}

\begin{small}
\begin{table*}[t]
\centering
\caption{Photometric and Spectroscopic properties \label{t:t1}}
\begin{tabular}{cccccccccccccc}
\hline
\hline
ID & $\alpha$ & $\delta$ & $z_{spec}$ & $i$ & $v-i$ & $i-z$ & $\log(r_e)$ & $\mu_e$ & S/N & $\sigma$ & $(\textrm{H}\gamma+\textrm{H}\delta)/2$ & [\ion{O}{2}] \\
 & $''$ & $''$  & & & & & ($kpc$) & & ($\textrm{\AA}^{-1})$ & ($km/s$) & (\AA) & (\AA) \\
\hline
19375 & 0&-50 & 1.089 & 21.72 & 1.71 & 1.04 & -0.410$\pm$0.012  & 21.75$\pm$0.05 & 26 & 198$\pm$25 & 4.1 &    -4.6 \\
19990 & -32&-35 & 0.964 & 21.32 & 1.89 & 1.00 & -0.388$\pm$0.007  & 21.45$\pm$0.03 & 49 & 159$\pm$14 & 2.5 & $>$-1 \\
20950 & -73&-6 & 0.964 & 21.18 & 2.07 & 1.07 &  0.0085$\pm$0.026  & 22.95$\pm$0.08 & 39 & 261$\pm$23 & $<$1 & $>$-1 \\
22432 & 83&41 & 1.135 & 22.38 & 2.00 & 1.42 & -0.109$\pm$0.040  & 23.35$\pm$0.13  & 21 & 217$\pm$20 & $<$1 &  -4.4 \\
21376 & 129&10 & 0.685 & 21.46 & 1.77 & 0.58 & -0.447$\pm$0.001  & 22.16$\pm$0.04 & 27 & 156$\pm$24 & --  &  -- \\
22239 & 236&35 & 0.660 & 20.67 & 1.67 & 0.50 & -0.842$\pm$0.002  & 19.83$\pm$0.03 & 40 & 177$\pm$19 & --  &  --\\
\hline
\tablecomments{Coordinates are in in arcseconds east and north of RA$=03^h32^m25^s$, Dec$=-27^{\circ}54'00''$. Errors in the magnitudes and colors are, respectively, 0.03 and 0.05 mag. Effective radii and surface brightnesses ($mag/arcsec^2$ at $r_e$) are measured in the $z$-band for objects 19375, 19990, 20950 and 22432, and in the $i$-band for objects 21376 and 22239. The listed errors in the velocity dispersions are fitting errors, and do not include a 5\% systematic error.}
\end{tabular}
\end{table*}
\end{small}

Stellar population models indicate that the low mass-to-light ratios
of the blue $z\approx 1$ galaxies may be due to an age difference of a 
factor of three. Alternatively, bursts involving 20-30 \% of the mass
can produce similar offsets.
The current sample is too small to determine the fraction of young
early type galaxies at $z \approx 1$ reliably.
Large,  mass selected samples are needed for this,
 as current samples are generally optically selected, and
therefore biased towards galaxies with lower mass-to-light ratios.

It is striking that the most massive galaxies have modest evolution in $M/L$, similar to what 
{van Dokkum} \& {Stanford} (2003)
found for massive 
cluster galaxies. The evolution of the galaxies with $M=6.07~r_e \sigma^2 \geq 3\times
10^{11} M_\odot$ 
({J\o rgensen} {et~al.} 1996)
in our sample is $\Delta \ln{M/L_B} = -1.17\pm0.14~z$.
The mass limit is comparable to the $M_*$ mass of an early type galaxy: if
we take the $\sigma_*$ of early type galaxies derived by 
{Kochanek} (1994)
of 225 km/s, we derive a typical mass of $M_*=3.1\times 10^{11} M_\odot$
based on the sample measured by 
{Faber} {et~al.} (1989). The sample as a whole evolves as 
$\Delta \ln{M/L_B} = -1.64\pm0.45~z$, whereas the sample with 
masses smaller than $3\times 10^{11} M_\odot$ evolves as 
$\Delta \ln{M/L_B} = -1.95\pm0.29~z$.


The results are therefore consistent with
little or no (recent) starformation in massive early type galaxies 
out to $z=1$,
and younger populations in less massive galaxies, possibly caused by
bursts
involving up to 30\% of the stellar mass.
Since these less massive galaxies have much more regular stellar 
populations at $z< 0.5$ without signs of recent star formation,
these
results are consistent with the downsizing seen in the field
population 
({Cowie} {et al.} 1996):
at progressively higher redshifts, more and more massive
galaxies are undergoing strong star formation.

It remains to be seen how this trend continues out to even higher
redshifts. The biases inherent in studies of galaxies at $z=2$ and
higher make it very hard to perform similar studies: the optical
light has shifted to the near-IR, and spectroscopy is extremely hard
at those wavelengths.

More studies at redshift $z\approx 1$ are needed to determine the
distribution of colors and mass-to-light ratios of the progenitors of
field early types. Such a determination should be based on
mass-selected
samples. Further studies of spectral energy distributions extending to
the restframe infrared will be very useful to constrain the star
formation
histories of the bluer galaxies better.
\acknowledgements
We thank the ESO staff for their support during the observations. We thank
C. Wolf for making available the COMBO17 catalogue.

\end{document}